\documentclass[conference,a4paper]{IEEEtran}
\IEEEoverridecommandlockouts

\usepackage{hyperref}
\usepackage[cmex10]{amsmath}
\usepackage{amssymb,amsfonts}
\usepackage{dblfloatfix}

\usepackage[ruled,vlined]{algorithm2e}
\usepackage{graphicx}
\graphicspath{{Figures/PDF/}{Figures/PNG/}}

\usepackage{booktabs}
\usepackage[numbers,compress]{natbib}
\usepackage{texnames}
\usepackage{bm,bbm}
\usepackage{orcidlink}

\usepackage{booktabs}
\usepackage{makecell}

\begin{document}

\title{Frame-Induced Doppler Self-Ambiguity in SiriusXM Satellite Signals for Passive Radar}

\author{
\IEEEauthorblockN{
	Khalid El-Darymli,~\IEEEmembership{SMIEEE},
	Christoph H.~Gierull,~\IEEEmembership{SMIEEE},
	Ryan A.~English, and
	Tony Laneve
}
\IEEEauthorblockA{
Department of National Defence, Defence Research and Development Canada, Ottawa, ON K1A 0Z4\\
\{khalid.el-darymli, christoph.gierull, ryan.english, tony.laneve\}@drdc-rddc.gc.ca
}
}

\maketitle

\begin{abstract}
Satellite Digital Audio Radio Service (SDARS) signals are attractive illuminators of opportunity for passive radar due to their continuous geostationary coverage and relatively high effective radiated power. However, the deterministic time-division multiplexed (TDM) framing employed by these waveforms introduces inherent self-ambiguity that can constrain coherent integration and detection performance. This paper analyzes the self-ambiguity function (SAF) of the XM/SiriusXM (SXM) waveform and explicitly links the observed Doppler periodicity to the repetition of the fast synchronization preamble (FSP) embedded in the TDM master frame. A frame-accurate XM TDM signal simulation is used to establish the expected ambiguity behavior, which is then validated experimentally using real direct-path data collected from the SXM-8 satellite. Results demonstrate pronounced and persistent Doppler replicas at integer multiples of the XM FSP repetition frequency, forming a deterministic Doppler ``comb'' (grating-lobe structure) that spans Doppler extents relevant to GEO-based passive radar operation with multi-second coherent processing intervals.
\end{abstract}

\vspace{-0.2cm}
\section{Introduction}\label{sec:intro}
\vspace{-0.18cm}
Passive radar systems exploit non-cooperative {illuminators of opportunity} and typically form range--Doppler products by correlating a surveillance channel with a reference (direct-path) channel over a coherent processing interval (CPI). Satellite Digital Audio Radio Service (SDARS) downlinks are attractive illuminators in this context~\cite{Provided_by_Christoph, 8038063, SABERTDA, Griffiths_NATO_report} due to continuous transmission from geostationary (GEO) platforms, wide-area coverage, and relatively high effective radiated power.

\begin{figure}[!t]
	\centering
	\includegraphics[width=\columnwidth]{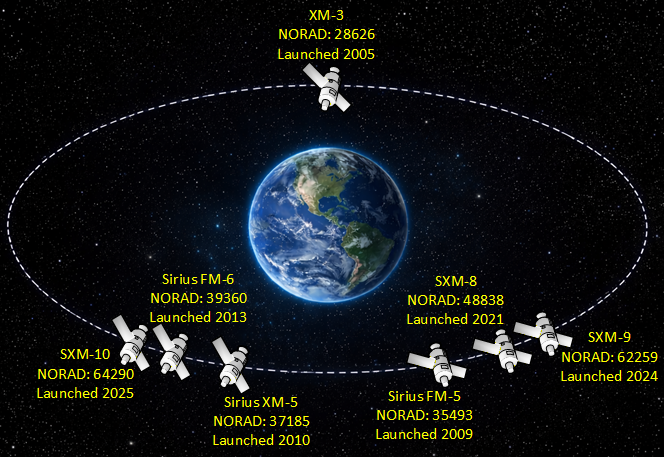}
	\caption{Operational SDARS satellites in geostationary orbit supporting XM and Sirius services. The SXM-8 satellite used in this study broadcasts the legacy XM-format waveform and is representative of current operational SDARS transmissions~\cite{keeptrack2025, SiriusXMWiki}.}
	\label{fig:geo_constellation}\vspace{-0.18cm}
\end{figure}

The self-ambiguity function (SAF) characterizes the intrinsic delay--Doppler response of a waveform and therefore governs the achievable sidelobe structure and masking behavior in correlation-based processing. Prior work has employed ambiguity analysis for bistatic and passive sensing (e.g., ocean remote sensing and surface scattering)~\cite{10487891, ShahGarrisonCorrelation, ShahGarrisonOcean}, often emphasizing noise-like waveforms, short CPIs, or moderate Doppler extents~\cite{GillGrenierChouinard}. In contrast, SDARS waveforms employ deterministic time-division multiplexed (TDM) framing designed for rapid receiver synchronization, introducing strong periodic structures that appear as discrete Doppler replicas (grating lobes) when multi-second coherent processing is used.

Fig.~\ref{fig:geo_constellation} illustrates the currently operational SDARS satellites in GEO orbit supporting the XM and Sirius services. In GEO bistatic configurations, long baselines and relative motion can give rise to extended Doppler extents, and practical passive radar processing often leverages long CPIs to improve sensitivity. Under these conditions, deterministic frame-induced Doppler replicas may dominate the range--Doppler map and can mask weak target returns unless explicitly accounted for. To the authors’ knowledge, the impact of deterministic SDARS framing on SAF behavior at extended Doppler offsets using real satellite data has not been previously reported.

The contributions of this paper are:
\begin{itemize}
\item An explicit analytical link between XM/SXM TDM framing and the expected Doppler replica spacing in the SAF.
\item Verification using a frame-accurate XM TDM signal simulation.
\item Experimental validation using real direct-path measurements from the SXM-8 satellite, demonstrating persistent frame-induced Doppler replicas across extended Doppler extents.
\end{itemize}

The remainder of this paper is organized as follows.
Section~\ref{sec:2} provides waveform context and key framing parameters.
Section~\ref{sec:3} describes the XM TDM master frame and derives the expected Doppler periodicity arising from FSP repetition.
Section~\ref{sec:4} summarizes the SAF formulation used throughout.
Section~\ref{sec:5} presents the frame-accurate simulation and its predicted ambiguity behavior.
Section~\ref{sec:6} validates the analysis using real direct-path data from SXM-8.
Sections~\ref{sec:7} and \ref{sec:8} discuss implications for passive radar processing and conclude the paper.

\section{SDARS Waveform Context}\label{sec:2}

\subsection{XM, Sirius, and SXM Overview}
North American SDARS service has historically been provided by XM and Sirius using distinct waveform structures. Following their merger, newer SiriusXM satellites were designed to preserve backward compatibility with legacy receivers. Consequently, operational SXM satellites continue to transmit the legacy XM waveform, including its modulation, framing, and synchronization characteristics. As a result, the XM TDM frame structure governs the signal behavior analyzed in this work.

\subsection{Key TDM Parameters}
Table~\ref{tab:tdm_params} summarizes key waveform and framing parameters for XM/SXM and Sirius signals. Sirius parameters are provided for contextual comparison.

\begin{table}[!t]
	\caption{Key SDARS TDM and Waveform Parameters~\cite{US8983364}}
	\label{tab:tdm_params}
	\centering
	\renewcommand{\arraystretch}{1.2}
	\begin{tabular}{l c c}
		\toprule
		\textbf{Parameter} & \textbf{XM / SXM} & \textbf{Sirius} \\
		\midrule
		Bit rate & 3.28 Mbps & 7.5168 Mbps \\
		Symbol rate & 1.64 Msps & 3.7584 Msps \\
		Fast sync preamble (FSP) length & 64 bits & 48 bits \\
		FSP frequency & 474.5 Hz & 3670 Hz \\
		Master frame preamble (MFP) & \makecell{128 bits\\(continuous)} & \makecell{255 bits\\(spread)} \\
		Master frame duration & 432 ms & 347 ms \\
		Padding (PAD) field & 192 bits & 0 bits \\
		RRC filter excess BW & 15\% & 20\% \\
		Total occupied BW & 1.886 MHz & 4.51008 MHz \\
		\bottomrule
	\end{tabular}\vspace{+0.3cm}
\end{table}

\begin{figure}[!t]
\centering
\includegraphics[width=\columnwidth]{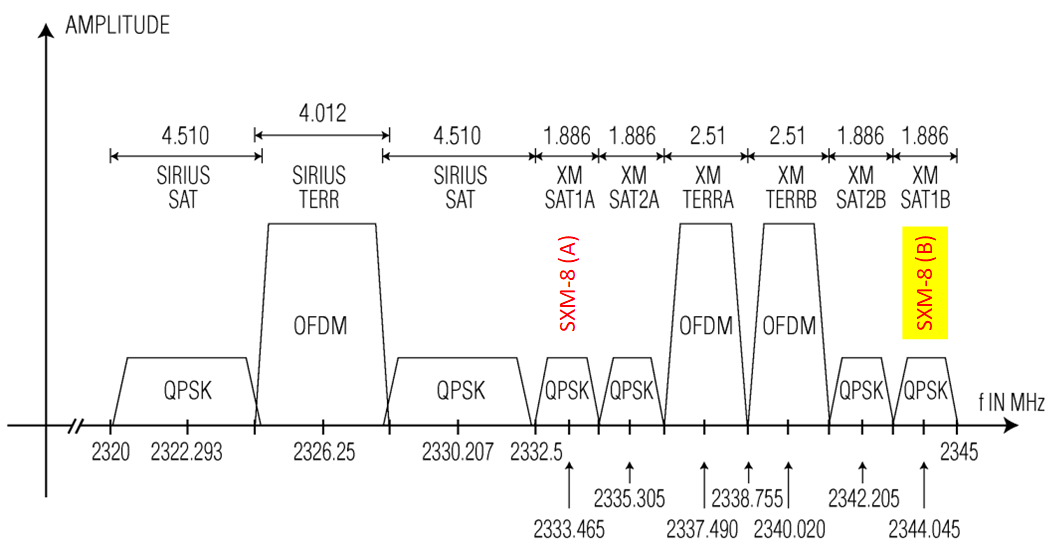}
\caption{SDARS S-band spectrum allocation for XM and Sirius systems~\cite{US8594559}.}
\label{fig:sdars_spectrum}\vspace{+0.3cm}
\end{figure}

For completeness, the Sirius waveform employs a distinct framing structure with a fast synchronization repetition frequency of  3670~Hz, implying Doppler replicas at substantially larger Doppler offsets than those associated with XM/SXM. Fig.~\ref{fig:sdars_spectrum} illustrates the SDARS S-band spectrum allocation; the SXM-8 (B) allocation used in this study is highlighted.

\section{XM TDM Frame Structure and Expected Doppler Periodicity}\label{sec:3}

\begin{figure}[!t]
\centering
\includegraphics[width=\columnwidth]{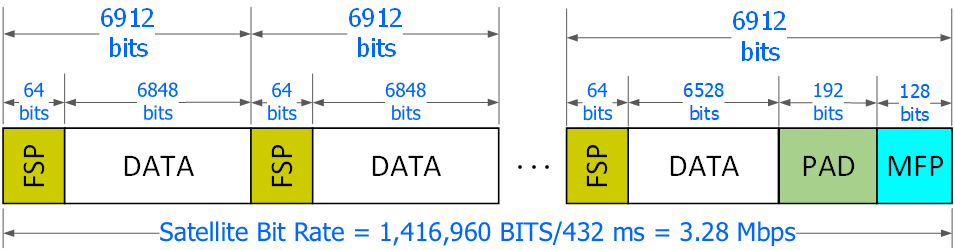}
\caption{XM TDM master frame structure illustrating the repeated FSPs that induce Doppler periodicity in the SAF (adapted from~\cite{US8983364}).}
\label{fig:xm_frame}
\end{figure}

The XM/SXM downlink employs a deterministic TDM waveform with a fixed master frame duration of $T_f=432$~ms. Each master frame contains 205 occurrences of a fast synchronization preamble (FSP), each 64 bits in length, interspersed with payload data and a short padding field. A schematic representation of the XM TDM master frame is shown in Fig.~\ref{fig:xm_frame}.
The corresponding FSP repetition frequency is
\begin{equation}
f_{\mathrm{FSP}}=\frac{205}{0.432}\approx 474.5~\text{Hz}.
\end{equation}

In ambiguity-function terms, this deterministic temporal repetition produces discrete Doppler replicas (a grating-lobe/comb structure) at
\begin{equation}
f_D=n f_{\mathrm{FSP}}, \quad n=\pm1,\pm2,\ldots
\end{equation}
independent of payload content. These replicas are an intrinsic waveform property; their visibility and masking impact increase with CPI length because coherent processing increases Doppler resolution and reinforces deterministic periodic components.

For convenience, the equivalent bistatic velocity associated with Doppler frequency $f_D$ is
\begin{equation}
v_D=f_D\frac{c}{f_c},
\label{eq:doppler_velocity}
\end{equation}
where $f_c$ is the carrier centre frequency and $c$ is the speed of light.

\section{Self-Ambiguity Function Framework}\label{sec:4}

For a complex baseband signal $s(t)$, the SAF is defined as~\cite{9100082,599331}
\begin{equation}
\chi(\tau,f_D)=\int_{-\infty}^{\infty} s(t)s^{*}(t-\tau)e^{-j2\pi f_D t}\,dt,
\end{equation}
where $\tau$ denotes delay and $f_D$ denotes Doppler frequency. In practice, the SAF is computed from sampled data over a finite CPI and visualized using its magnitude. For periodically framed waveforms such as XM/SXM, deterministic temporal repetition leads to discrete Doppler replicas that can dominate the ambiguity surface under long coherent integration.

\section{XM TDM Signal Simulation}\label{sec:5}

The XM waveform is proprietary. To support transparent and reproducible analysis, a frame-accurate simulator was developed using publicly available information from SDARS receiver patent literature and publicly available decoder documentation~\cite{US8983364,US8594559,stmicro_sta400a_2025}. The simulator implements the publicly documented elements that govern ambiguity behavior, including TDM framing, channel coding, interleaving, QPSK modulation, and root-raised-cosine filtering. This approach reproduces the deterministic frame-induced structure relevant to SAF analysis while not claiming access to proprietary implementation details.

\begin{figure}[!t]
\centering
\includegraphics[width=245pt]{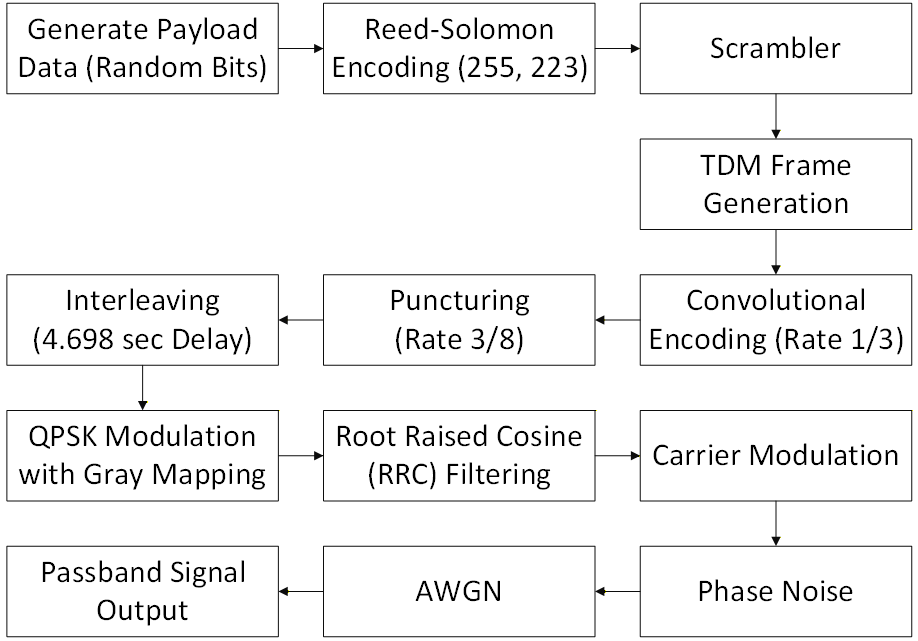}
\caption{Block diagram of the frame-accurate XM TDM signal simulation.}
\label{fig:sim_block}
\end{figure}

\begin{figure}[!t]
\centering
\includegraphics[width=\columnwidth]{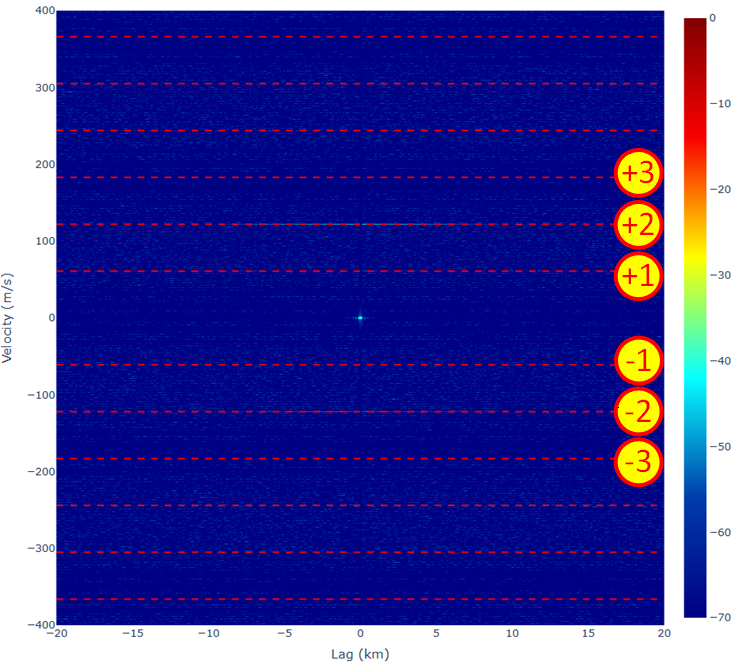}
\caption{Simulated SAF of the XM/SXM waveform using a CPI of 3~s. Dashed red lines indicate the expected Doppler-replica (harmonic) locations in the SAF, as given by Eq.~\eqref{dashed-line_v}.}
\label{fig:sim_saf}
\end{figure}

\begin{figure}[!t]
\centering
\includegraphics[width=\columnwidth]{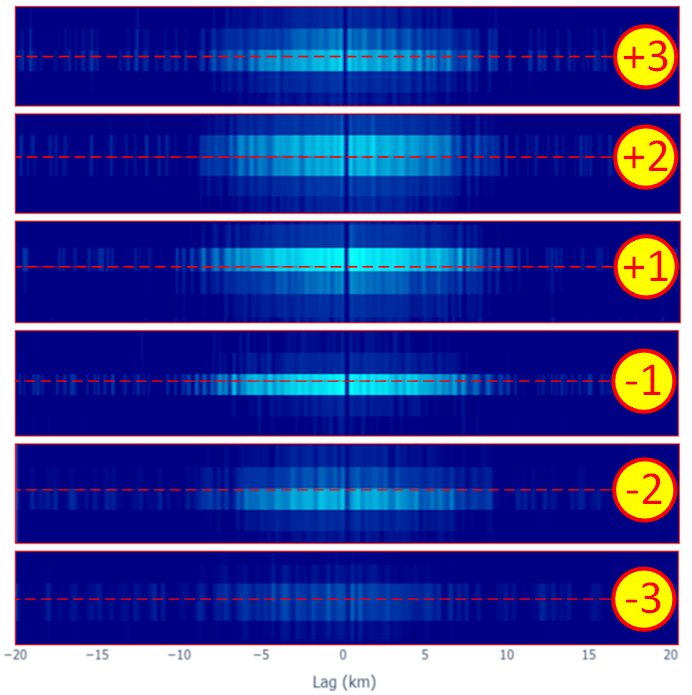}
\caption{Zoomed-in views of the simulated SAF in Fig.~\ref{fig:sim_saf} highlighting the deterministic Doppler replica (comb) structure induced by XM/SXM framing. The dashed red lines demonstrate close agreement between theory and simulation.}
\label{fig:sim_saf_zoom}
\end{figure}

The simulation focuses on the SXM-8 (B) carrier centred at $f_c=2344.045$~MHz (Fig.~\ref{fig:sdars_spectrum}). A CPI of 3~s was used for SAF computation. Using Eq.~\eqref{eq:doppler_velocity} and $f_{\mathrm{FSP}}\approx 474.5$~Hz, the expected Doppler-replica velocity spacing is approximately 60.7~m/s, with harmonics at
\begin{equation}\label{dashed-line_v}
 v_D=\pm60.7, \pm121.5, \pm182.2, \pm242.9,\pm303.6, ...\,\text{[m/s]}
\end{equation}
A block diagram of the simulation processing chain is shown in Fig.~\ref{fig:sim_block}.
The simulated SAF is shown in Fig.~\ref{fig:sim_saf}. The dashed red lines indicate the predicted Doppler replica locations given by Eq.~\eqref{dashed-line_v}. Fig.~\ref{fig:sim_saf_zoom} provides zoomed-in views around selected harmonics, highlighting the grating-lobe/comb structure induced by deterministic frame repetition and confirming agreement between the predicted and observed replica locations.

\section{Experimental Data and SAF Results}\label{sec:6}
The SXM-8 (B) direct-path signal was recorded on June~21,~2023 using a Cobham flat-panel left-hand circularly polarized (LHCP) antenna. Range--Doppler and SAF products were computed using a CPI of 3~s, and the final measured result was formed via incoherent summation over approximately 2~minutes of recorded data to stabilize the displayed ambiguity structure.

\begin{figure}[!t]
	\vspace{+0.3cm}
	\centering
	\includegraphics[width=\columnwidth]{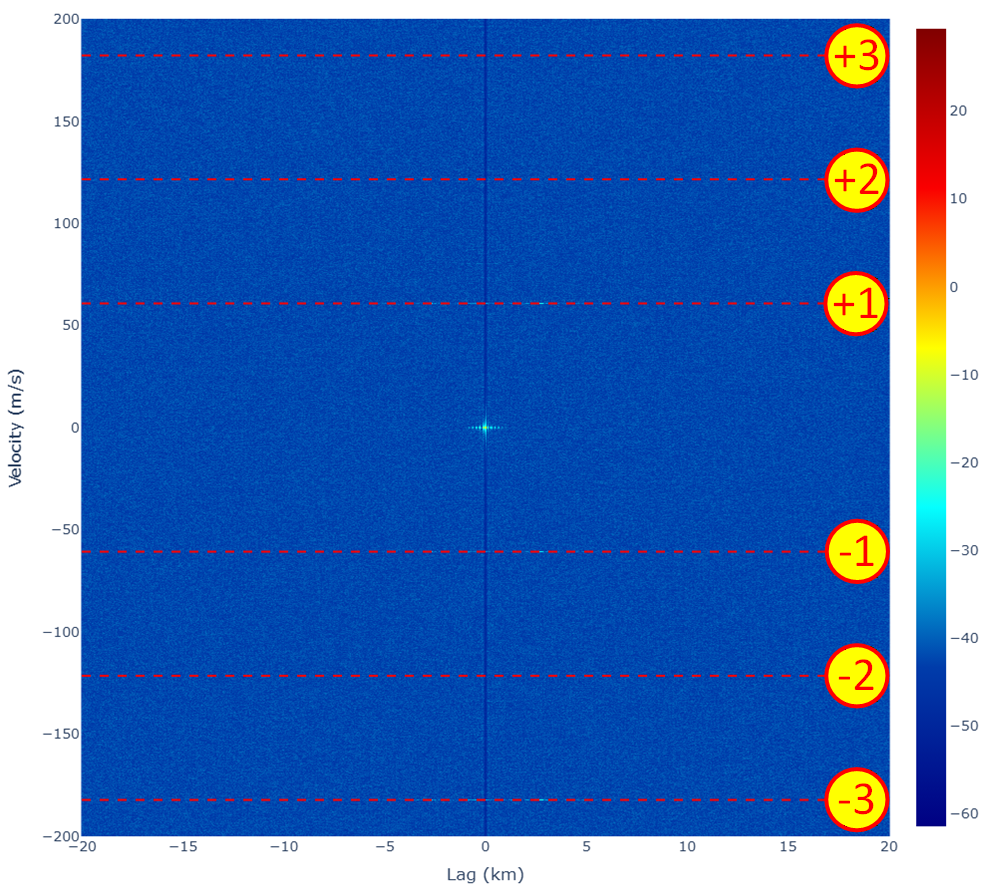}
	\caption{Measured SAF of the SXM-8 (B) direct-path signal. Dashed red lines indicate the expected Doppler-replica (harmonic) locations in the SAF, as given by Eq.~\eqref{dashed-line_v}. The plot corresponds to an incoherent sum over approximately 2~minutes of recorded data, using a CPI of 3~s for each range--Doppler computation.}\vspace{+0.02cm}
	\label{fig:real_saf}
\end{figure}

The measured SAF is shown in Fig.~\ref{fig:real_saf}. Pronounced Doppler replicas are observed at spacings consistent with the XM FSP repetition frequency, matching the simulated and predicted locations. Fig.~\ref{fig:real_saf_zoom} presents zoomed-in views around selected Doppler harmonics, confirming close agreement between prediction and measurement.

All simulations and data processing were performed on a Microsoft Azure compute environment equipped with NVIDIA A100 (80~GB) PCIe GPUs.

\section{Discussion}\label{sec:7}
The results demonstrate that frame-induced Doppler replicas are an intrinsic characteristic of the XM/SXM waveform, arising from deterministic repetition of the FSP embedded in the TDM frame. This mechanism produces a stable Doppler ``comb'' at integer multiples of the FSP repetition frequency, independent of payload content, and becomes increasingly prominent as coherent processing intervals are extended.

\begin{figure}[!t]
	\vspace{+0.37cm}
	\centering
	\includegraphics[width=250pt]{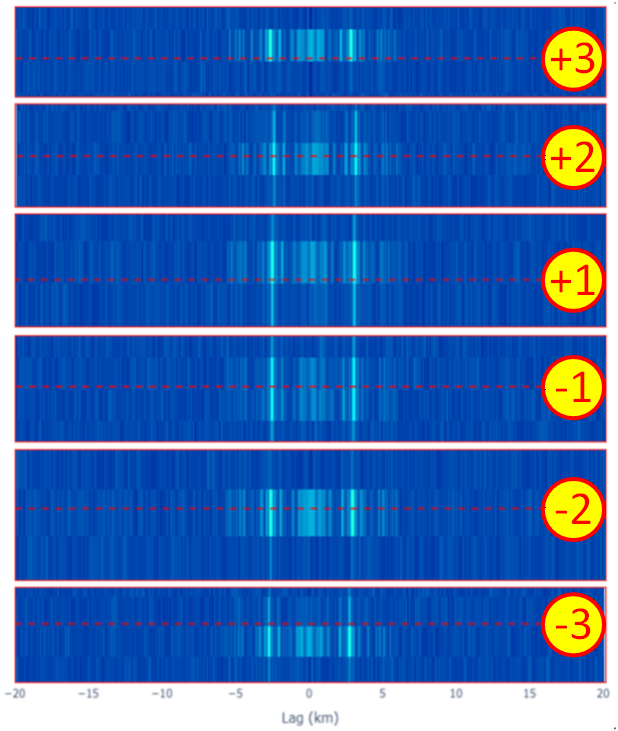}
	\caption{Zoomed-in views of the measured SAF in Fig.~\ref{fig:real_saf} around selected Doppler replicas. The dashed red lines demonstrate close agreement between theory and experiment.}
	\label{fig:real_saf_zoom}
\end{figure}

Measured self-ambiguity results derived from the reference (direct-path) channel exhibit pronounced ambiguity energy centred at a non-zero delay corresponding to $\pm2.736$~km (Fig.~\ref{fig:real_saf_zoom}), with peak levels on the order of $-22.8$~dB relative to the main response and well above the noise floor ($\approx-60$~dB). This delay offset has been observed consistently across multiple experimental datasets acquired with SXM signals; however, its precise origin remains uncertain.

The XM/SXM waveform is proprietary, and the frame-accurate simulation employed in this study is necessarily limited to publicly available patent literature and decoder documentation. As such, the simulator captures the known framing, synchronization, and modulation structures that govern Doppler-domain ambiguity behavior, but cannot represent any undocumented or proprietary signal features present in operational transmissions. Importantly, the observed delay offset does not affect the predicted or measured Doppler replica spacing, which remains in close agreement between theory, simulation, and experiment.

From a passive radar perspective, the resulting Doppler comb acts as a deterministic ambiguity pedestal that can elevate sidelobe levels and mask weak target responses at Doppler offsets coincident with the replica harmonics. This motivates frame-aware processing strategies, including informed CPI selection, Doppler-domain mitigation at known harmonic locations, and reference-channel conditioning that explicitly accounts for repeated synchronization structures. Future work will quantify detection and false-alarm impacts in representative GEO bistatic configurations and assess practical mitigation approaches for SDARS-based passive radar.

\vspace{+0.3cm}
\section{Conclusion}\label{sec:8}

This paper analyzed the SAF of the XM/SXM SDARS waveform with emphasis on deterministic frame-induced Doppler periodicity. By linking the XM TDM structure to an expected Doppler replica spacing, close agreement was demonstrated between theory, frame-accurate simulation, and real direct-path measurements using SXM-8 (B) data acquired on June~21,~2023. The results highlight the need for frame-aware passive radar processing when exploiting SDARS illuminators of opportunity and provide a foundation for ambiguity-mitigation and target-detection studies. Extensive real-world passive radar experiments with SXM have been conducted at DRDC-Ottawa Research Centre over multiple years and will be presented in future publications.

\setlength{\bibsep}{6pt}
\small
\bibliographystyle{IEEEtranN}
\bibliography{references}

\end{document}